\documentclass[twoside,leqno,twocolumn]{article}

\usepackage[letterpaper]{geometry}
\usepackage{siamproceedings}

\usepackage[T1]{fontenc}
\usepackage{amsfonts}
\usepackage{graphicx}
\usepackage{booktabs}
\usepackage{enumitem}
\usepackage{xcolor}
\usepackage{cuted}
\usepackage{capt-of}
\usepackage{tikz}
\usetikzlibrary{arrows.meta,decorations.pathreplacing,fit,shapes.geometric,calc}

\newcommand{\calibLo}{100}            %
\newcommand{\calibHi}{200}            %
\newcommand{\numOptimizers}{256}      %
\newcommand{\numParallelOptimizers}{229}%
\newcommand{\numRoster}{twenty}       %
\newcommand{\screenBudget}{15{,}000}  %
\newcommand{\screenRuns}{2}           %
\newcommand{\escalBudget}{100{,}000}  %
\newcommand{\escalRuns}{5}            %
\newcommand{\masterSeed}{20260601}    %
\newcommand{\timecap}{60}             %

\newcommand{\SATevolvedTime}{12.0}    %
\newcommand{\SATevolvedTimeKnob}{11.1}%
\newcommand{\SATvsRandomPT}{13}       %
\newcommand{\SATvsRandomSearch}{26}   %
\newcommand{\SATbestOpt}{MultiScaleCMA}
\newcommand{\SATrandomControl}{0.47}  %
\newcommand{\SATtypical}{0.01}        %
\newcommand{\SATvsTypical}{1{,}000}   %
\newcommand{\SATcompEasier}{387}      %
\newcommand{\SATcompEasierPct}{78}    %

\newcommand{\SATdpllChamp}{1.36}      %
\newcommand{\SATdpllKissat}{2.1}      %
\newcommand{\SATdpllCadical}{3.4}     %
\newcommand{\SATdpllCMS}{13.1}        %
\newcommand{\SATdpllMarch}{1.36}      %
\newcommand{\SATdpllSlowest}{1.1}     %
\newcommand{\SATdpllVsTypical}{130}   %

\newcommand{\TSPevolvedTime}{17.3}    %
\newcommand{\TSPevolvedHeldout}{12.9} %
\newcommand{\TSPevolvedTimeB}{13.5}   %
\newcommand{\TSPevolvedHeldoutB}{7.99}%
\newcommand{\TSPlibHardest}{2.34}     %
\newcommand{\TSPlibHardestHeldout}{2.24}%
\newcommand{\TSPShiwaVsTSPLibHeldout}{3.6}%
\newcommand{\TSPDiagVsTSPLibHeldout}{5.7}%
\newcommand{\TSPbestOpt}{DiagonalCMA}
\newcommand{\TSPrandomControlTrain}{0.70}%
\newcommand{\TSPrandomControlHeldout}{0.79}%

\newcommand{\HAMevolvedTime}{1.12}    %
\newcommand{\HAMevolvedHeldout}{0.48} %
\newcommand{\HAMnullTrain}{0.26}      %
\newcommand{\HAMnullHeldout}{0.17}    %
\newcommand{\HAMfhcpSolved}{7.88}     %
\newcommand{\HAMbestOpt}{DiscreteBSO}
\newcommand{\HAMrandomControlTrain}{0.35}%
\newcommand{\HAMrandomControlHeldout}{0.28}%
\newcommand{\HAMvsRandomTrain}{3.2}   %
\newcommand{\HAMvsRandomHeldout}{1.8} %

\newcommand{\FSSPsatChamp}{18.2}      %
\newcommand{\FSSPsatChampHeld}{16.3}  %
\newcommand{\FSSPsatChampTransfer}{0.33}%
\newcommand{\FSSPoptChamp}{0.54}      %
\newcommand{\FSSPoptChampHeld}{0.52}  %
\newcommand{\FSSPbigChamp}{7.24}      %
\newcommand{\FSSPbigChampHeld}{7.03}  %
\newcommand{\FSSPbigVsHardestRandom}{11}%
\newcommand{\FSSPbestOpt}{DiagonalCMA}%
\newcommand{\FSSPrandomControl}{0.67} %
\newcommand{\FSSPrandomControlHeld}{0.65}%
\newcommand{\FSSPbigTypical}{0.027}   %
\newcommand{\FSSPbigVsTypical}{250}   %
\newcommand{\FSSPlargeChamp}{238.6}   %
\newcommand{\FSSPlargeChampHeld}{235.3}%
\newcommand{\FSSPlargeRandom}{4.50}   %
\newcommand{\FSSPlargeTypical}{0.13}  %
\newcommand{\FSSPlargeVsTypical}{1{,}700}%
\newcommand{\FSSPlargeVsRandom}{53}   %
\newcommand{\FSSPlargeBudget}{50{,}000}%
\newcommand{\FSSPlargeRuns}{2}        %
\newcommand{\FSSPlargeCap}{600}       %
\newcommand{\FSSPtaillardHardest}{4.5}%
\newcommand{\FSSPlargeVsTaillard}{53} %
\newcommand{\FSSPvrfHardest}{3.8}     %
\newcommand{\FSSPlargeVsVRF}{62}      %
\newcommand{\ortoolsVersion}{9.15}

\newif\iffullversion
\fullversiontrue %
\newcommand{\appref}[1]{\iffullversion\cref{#1}\else see the full version\fi}
\newcommand{\appin}[1]{\iffullversion\cref{#1}\else the full version\fi}

\begin{document}

\iffullversion
\newcommand\relatedversion{\thanks{Full version of a paper in the
proceedings of ALENEX 2027.}}
\else
\newcommand\relatedversion{\thanks{Full version: \protect\url{https://arxiv.org/abs/XXXX.XXXXX}}}
\fi

\title{\Large Hard Instances on Demand\thanks{All authors were supported by
the Czech Science Foundation (GA \v{C}R), project 25-17221S.
M.~Kouteck\'y was additionally partially supported by Charles University
project UNCE~24/SCI/008 and by the ERC-CZ project LL2406 of the Ministry of
Education, Youth and Sports of the Czech Republic.}\relatedversion}

\author{Jamie Fravel\thanks{Computer Science Institute, Faculty of Mathematics
and Physics, Charles University, Prague, Czech Republic.
\texttt{jfravel@iuuk.mff.cuni.cz}, \texttt{koutecky@iuuk.mff.cuni.cz},
\texttt{tsitseip@iuuk.mff.cuni.cz}.}
\and Martin Kouteck\'y\footnotemark[2]
\and Pavlo Tsitsei\footnotemark[2]}

\date{}

\onecolumn
\maketitle

\iffullversion\else
\fancyfoot[R]{\scriptsize{Copyright \textcopyright\ 2027 by SIAM\\
Unauthorized reproduction of this article is prohibited}}
\fi

\begin{abstract}
The conclusions of an experimental study of algorithms are, in many ways, only as good as the instances they are run on. For well-studied problems, benchmark libraries have taken decades
to curate while evaluations of new problems often fall back on random instances due to lack of real-world data.
However, random instances are known to be misleading---they may say
more about the structure of random generation than about the
algorithmic question at hand.

We treat instance generation as an optimization
problem: parameterize the instance space, take the solver's running time
as the objective, and let a black-box optimizer search for hard instances.
We develop a widely applicable workflow built on the Nevergrad optimization
platform and demonstrate it on four NP-hard problems:
satisfiability, traveling salesperson, Hamiltonian cycle, and
non-permutation flow-shop scheduling.
We propose this as a routine component of
experimental papers; the search needs only the solver, an instance
encoder, and a compute budget. The four problems---chosen for their
strong external baselines---validate the effectiveness of our scheme. Our evolved instances
beat the hardest instance of a budget-matched, random draw in all four case studies and beat them by an order of magnitude or more in three cases.  They are often competitive
with long-curated benchmark sets: our SAT champion is harder (for a
three-solver portfolio) than \SATcompEasierPct\% of the 498 SAT
Competition formulas in its size class, matched or exceeded only by
families that encode expert insight into hardness; our 20-city TSP
champions take several times longer on held-out solver seeds than the
hardest measured classical TSPLIB instance (which has 229 cities); and our
non-permutation flow-shop champion takes \FSSPlargeChamp~seconds of
\textsf{CP~Optimizer} time to solve to proven optimality (compared to 4.5~seconds for the
hardest of 100{,}000 random draws) while also denying \textsf{CP-SAT} an
optimality proof. The optimizer even recovers known structural characteristics from running
time alone: it locates a SAT clause/variable ratio near the phase
transition and, attacking the two scheduling engines separately, finds
structurally different champions for each.
\end{abstract}

\clearpage
\twocolumn

\section{Introduction}
\label{sec:intro}

Good test instances can be scarce. For a handful of classical
problems, decades of community effort have produced benchmark libraries:
the SAT competitions~\cite{JarvisaloLBS12,FroleyksHIJS21},
TSPLIB~\cite{Reinelt91},
MIPLIB~\cite{GleixnerEtAl21}, and the Taillard instances for shop
scheduling~\cite{Taillard93} are all good examples. Each of these serves an NP-hard problem whose
solvers are \emph{exact} (i.e., they prove the answers they return) and, even
there, curation is a running concern: the SAT
competitions solicit new benchmarks every year and MIPLIB is in its sixth edition, with both collections requiring re-curation as solvers improve and once-challenging instances become easy.

The situation is worse away from the classics. When Hiermann et
al.~\cite{HiermannPRH16} introduced the electric fleet-size-and-mix
vehicle-routing problem with time windows and recharging stations
(by now a classical problem warranting reviews of its own~\cite{KucukogluDC21}), they also had to
create its benchmark set. A new problem may have no established
instance corpus while a new algorithm for a known problem may face only a thin
one. The common fallback is
random instance generation, which has well documented dangers.
Johnson~\cite{Johnson02}, in his classic essay on experimental
algorithmics, lists concentration on unstructured
random instances among his named pet peeves, writing that such instances
``may actively mislead us as to the difficulty of the problem.'' Johnson's example
is random asymmetric TSP distance matrices, which become easy as the
instances grow (for the commonly used bounded coefficient ranges).
Hooker~\cite{Hooker95} and McGeoch~\cite{McGeoch12} likewise make
test-problem choice central to valid
algorithm experiments. Many natural random
distributions produce systematically easy instances; for random 3-SAT, the classic observation is that hardness concentrates near the
satisfiability threshold~\cite{CheesemanKT91,SelmanML96}, a picture
since revisited critically~\cite{SleegersOVB20}. Our own measurements agree that a positive evaluation on
such instances says little: in our SAT and TSP experiments,
even the hardest instance of a budget-matched, random draw is an order
of magnitude easier than the evolved champions (\cref{sec:results}). On the other hand, a new algorithm performing well on a thin, albeit established, benchmark set may not imply good performance in general, merely good tuning to the benchmark.

\paragraph{Adversarial instance generation.}
We propose to attack both problems the same way: search for hard
instances directly. We parameterize a space of instances of a given size
and let a black-box optimizer maximize the solver's measured running
time over it. The optimizer plays the role of adversary and needs nothing
apart from the ability to run the solver. Hardness found this way is necessarily relative to the
attacked solver (\cref{sec:prelims}). Hardness for a \emph{portfolio}
of solvers is a stronger property, and our approach finds that as well
(\cref{sec:results:sat}).

\paragraph{A new default for experimental papers.}
Mu{\~n}oz and Smith-Miles~\cite{MunozSM20} call for test instances that
``expose the strengths and weaknesses of algorithms and drive algorithm
development''. Our central proposal is a
concrete \emph{default} toward that goal: alongside performance on standard
benchmarks, an experimental paper should also report the hardest instances
a fixed-budget adversarial search finds against its own solver.
Our paper gives a concrete, easy-to-follow workflow which only requires the solver, an instance encoder, and an evaluation budget
(\cref{sec:discussion} discusses the cost), ingredients every
experimental paper already has.
Adversarial testing tells the reader what benchmark averages cannot: where the solver breaks, and how badly.

\paragraph{Validation: why these four problems.}
The obstacles to this proposal were never conceptual. A
platform like Nevergrad plausibly supplies the adversary, but it
offers several hundred optimizers with no guidance on which of them
can attack a solver's running time. More fundamentally, beating random
sampling is not by itself evidence of genuine hardness.
Whether the found instances are
impressive or trivial can only be judged against strong existing
benchmarks, which new problems (those that most require this sort of analysis) do not
have. We therefore validate the workflow on four problems chosen for
their existing, external baselines.
Boolean satisfiability (SAT) and the traveling salesperson problem
(TSP) have decades of crafted adversarial families (pigeonhole,
Tseitin, van der Waerden formulas, Hard-TSPLIB). Hamiltonian cycle
(HAM) has a curated challenge set (FHCP), though at larger sizes than
ours. Flow shop (studied here in its general non-permutation form
(NPFS)) has the Taillard~\cite{Taillard93} and VRF~\cite{ValladaRF15}
suites created for the restricted permutation variant, but the
standard test bed for NPFS as well~\cite{BenavidesR16,BenavidesR18}. Performing credibly
against these benchmarks licenses our recommendation to use the
workflow herein where it is needed most: on problems whose only baseline is
random.

The four problems also differ in their input data. SAT and HAM
have no numbers at all---a formula is clauses of literals, a graph is
edges, and HAM is the unweighted feasibility counterpart of TSP.
TSP and NPFS have integer weights whose allowed range $U$ lets
the search probe a solver's sensitivity to coefficient magnitude. 
Our NPFS experiments use $U=100$ derived from the scale of Taillard's generator, and TSP reports $U=10^3$ and $U=10^4$ as separate experiments.
We fix a relatively low value of $U$ so that coefficient scale alone (through, for example, numerical instability) does not drive the hardness.

\paragraph{Objectives beyond running time.}
Our method can be used for other optimization objectives beyond running time, as long as it is a reproducibly measurable scalar. For example, the number of conflicts \textsf{MiniSat} reports (a machine-independent proxy) or the length of the proof
certificate \textsf{kissat}~\cite{BiereFFH20} emits. One can also invert the goal to search for instances inside a
prescribed difficulty band to build a graded benchmark suite, or for
instances which fill the gaps in an
existing corpus~\cite{SmithMilesL12}.

\paragraph{Contributions.}
    \begin{itemize}[nosep]
    \item A proposal for experimental practice: report, alongside
    standard benchmarks, the hardest instances a fixed-budget adversarial
    search finds against your solver. 
    
    \item To make this affordable, we designed a common workflow
    (\cref{sec:protocol}): calibrate the instance size, screen a fixed
    roster of \numRoster{} black-box optimizers (distilled from our
    one-time \numOptimizers-optimizer screen) at a modest budget, then
    escalate the budget for the few optimizers the screening singles
    out.
    
    \item An instantiation of the above workflow for SAT, TSP, HAM, and NPFS,
    including the instance parameterizations which are the main
    problem-specific ingredient (\cref{sec:prelims,sec:problems}).
    
    \item An experimental comparison against random instances, curated
    benchmark libraries, and hand-crafted adversarial families
    (\cref{sec:results}). 
        \begin{itemize}[nosep]
        \item For SAT, the evolved champion is \SATvsRandomSearch$\times$ harder than equal-budget random sampling for a three-solver portfolio. Only families built on expert hardness insight exceed it in the size-matched SAT Competition archive. 
        \item For TSP, the two headline champions are several times harder than the hardest of the measured classical TSPLIB instances (up to 300 cities) while having only one-tenth of its cities. This result is measured on held-out solver seeds to account for possible optimizer over-fitting.
        \item For flow shop, the $10\times5$ champion takes \FSSPlargeChamp{} seconds of \textsf{CP~Optimizer} time to solve to proven optimality. This is a factor of \FSSPlargeVsRandom{} above the hardest of $100{,}000$ random draws, of \FSSPlargeVsVRF{} above the hardest official VRF instance of identical size, and of \FSSPlargeVsTaillard{} above the hardest official Taillard instance of the size-closest group ($20\times5$), even though it has twice as many jobs. Moreover, here the hardness transfers between solvers: the $8\times4$ champion bred against \textsf{CP~Optimizer} exceeds \textsf{CP-SAT}'s 60-second cap.
        \item For HAM, the gain over budget-matched random search is a factor of two to three, and we find that the optimizer discovers structure. The champions it produces are sparse, asymmetric, and barely non-Hamiltonian, sitting just past the Hamiltonicity boundary.
        \end{itemize}
    \end{itemize}

\subsection{Related work}
\label{sec:related}

Van Hemert~\cite{vanHemert06} evolved CSP, SAT, and TSP instances to
maximize solver effort,
finding harder instances than popular benchmarks. His TSP instances target
Lin--Kernighan-style local search, and Julstrom~\cite{Julstrom09}
generalized the idea. Hougardy and Zhong~\cite{HougardyZ21} hand-designed
a Euclidean
TSP family over a million times slower for \textsf{Concorde} than similarly sized
TSPLIB instances. Fischer et al.~\cite{FischerSHM05} relate structure
and hardness from the opposite direction: starting from highly
structured, polynomially solvable TSP instances, they remove cities or
perturb coordinates and measure how the hardness changes for local
search and branch-and-cut. Sleegers and Van den Berg~\cite{SleegersB22} evolved
14-vertex graphs
requiring about 28$\times$ as much search by a Hamiltonian-cycle
backtracker---not \textsf{Concorde}---as their hardest Hamiltonian
graph. They later exhibited instances which are hard for every solver they
tested~\cite{SleegersTB22}; our portfolio objective guarantees
hardness only for the portfolio's members. Goerigk and Maher~\cite{GoerigkM20} use optimization
models to generate
robust-optimization instances up to 500$\times$
harder than random ones. This work is extended by Goerigk and
Khosravi~\cite{GoerigkK24}. Later adversarial work
includes SAT and TSP, but targets solver differences, failures, or
learned solvers rather than releasing instances evolved against an
exact solver's measured running time~\cite{LiuTY22,
ScottMG20,HuZZZ26,GeislerSSBG22,LuEtAl23,NikoleitANR26}.
HardSATGen~\cite{LiEtAl23} and HardCore~\cite{CotnareanuEtAl24}
generate variants of already-hard SAT formulas and test whether the variants
retain their difficulty. They do not search for instances harder than their
seeds. The scheduling benchmark suites
(Taillard, VRF, Demirkol) are introduced with the flow-shop results
in \cref{sec:results:jssp}.

Declarative generator models can express structured instance domains that are
awkward for direct box encodings like ours~\cite{AkgunEtAl19,DangEtAl22}.
These systems use \texttt{irace} (a package implementing the
``iterated racing'' method) to tune a small generator parameterization
toward a solving-time band or a solver separation, rather
than directly maximizing one instance's runtime~\cite{AkgunEtAl19,AkgunEtAl20,DangEtAl22}.
For our fixed-size SAT domain, neither obvious parameterization fits:
automatically derived parameters are too coarse to tune
hardness-relevant correlations, whereas exposing the literals directly
exceeds stock \texttt{irace}'s 1,024-dimensional Sobol initializer.
The published experiments also provide no random or established-hard
baseline directly comparable to ours.

Instance space analysis diagnoses gaps and solver-specific regions in an
existing corpus~\cite{SmithMilesL12,SmithMilesM23}. Algorithm configuration
applies the same outer-loop pattern in the opposite direction.
SMAC~\cite{HutterHL11} and \texttt{irace}~\cite{LopezIbanezEtAl16}
tune a solver to run fast on fixed instances;
we hold the solver fixed and search for slow instances.

\section{Method}
\label{sec:prelims}

\subsection{Hardness is relative to a solver}
An \emph{instance} of a computational problem has no running time by
itself; only a solver applied to it does. Indeed, in theory there is no
such thing as a hard instance at all: for any single instance there is
an algorithm that stores its answer and returns it in constant time.
We therefore define the
\emph{hardness} of an instance $I$ with respect to a solver $S$ as the
CPU time, in seconds, that $S$ takes to solve $I$ (to proven
optimality, when $S$ is an optimizing solver) on a fixed reference
machine. When
something closer to hardness for a \emph{problem} is wanted and several
solvers $S_1,\dots,S_k$ are available, we use the portfolio hardness
$\min_i \operatorname{time}(S_i, I)$: an instance counts as hard only
if every solver in the portfolio struggles with it. Our SAT
experiments use exactly this objective (\cref{sec:results:sat}). Repeated measurements and randomized solver
seeds are aggregated by the problem-specific protocols in
\cref{sec:setup}; the result tables report medians over the stated seed
sets or repetition counts after a final clean re-measurement.

\subsection{Instance generation as black-box optimization}
Fix a problem, an instance size, and a solver (or portfolio). An
\emph{encoding} is a box
$[l,u]^d = \{x \in \mathbb{R}^d \mid l \le x_i \le u \text{ for each } i\}$,
given by bounds $l<u$ and a dimension $d$ (or its integer analogue in
$\mathbb{Z}^d$), together with a \emph{decoder} that turns each
point of the box into an instance of the chosen size. The objective function of a
point is (a decreasing transform of) the measured hardness of the decoded
instance. Any derivative-free optimizer can now search the
box:\footnote{The box is the simplest search space, not the limit:
Nevergrad also optimizes over integer arrays, log-scaled scalars,
categorical choices, and nested tuples. Hard constraints are the real
limitation; feasibility is best built into the decoder, as our
score-and-threshold HAM encoding does.}
such an optimizer needs no gradients and no view into the
objective. In an \emph{ask/tell} loop it proposes a point (ask),
receives the measured loss (tell), and biases future proposals toward
better regions. Methods differ in how: by evolving a population,
adapting a sampling distribution, or fitting a surrogate.
The encoding is the main problem-specific ingredient of the approach,
and we deliberately use naive encodings to test whether hard instances
can be found without domain insight.

We run the optimizers of the Nevergrad platform~\cite{BennetDMRTT21},
which
offers several hundred derivative-free optimizers behind one ask/tell
interface, including the CMA-ES family, particle swarm variants,
differential evolution, bandit-style discrete methods, and
``wizard'' portfolios such as \textsf{NGOpt} and \textsf{Shiwa} that select among the
others. How each individual optimizer works is beyond the scope of
this paper: the workflow treats them as interchangeable adversaries,
and \cref{sec:protocol} reduces the several hundred to a fixed roster
of \numRoster{}.

All experiments run through our
experiment-management layer around Nevergrad, released as part of the
artifact. A problem plugs in by implementing an
encode/decode/solve interface, and the layer manages the configured
optimizer campaigns, parallel execution, per-candidate aggregation,
and checkpointed result archives (artifact link in \cref{sec:setup}).

\subsection{Common workflow}
\label{sec:protocol}

The workflow has three default stages: calibrate the instance size, screen a
fixed roster of optimizers at a modest budget, and escalate the budget
for the optimizer families that the screening singles out. The solver,
encoding, aggregation rule, and occasionally the budget adapt to the case
study; \cref{sec:setup} records those deviations.

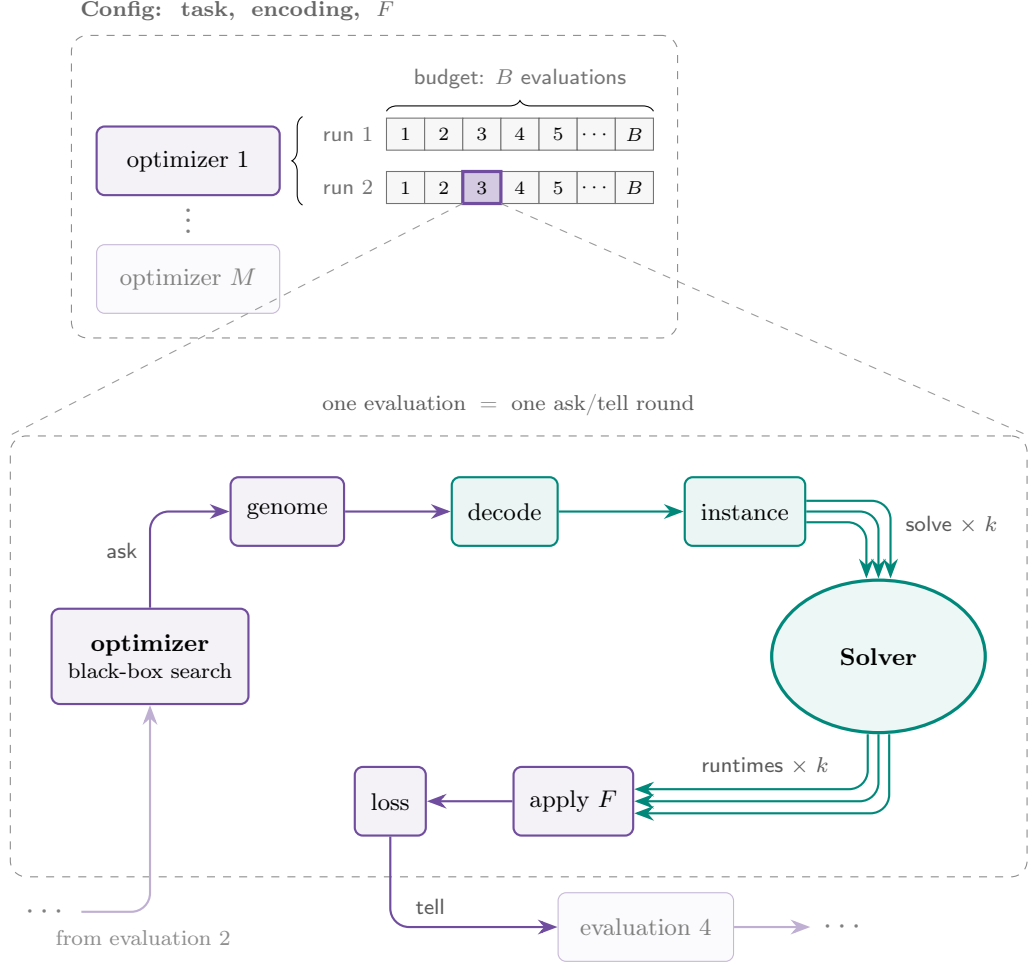
\begin{figure*}[t!]
  \centering
  \resizebox{0.80\textwidth}{!}{\definecolor{optpurple}{RGB}{112,77,158}
\definecolor{evalteal}{RGB}{0,137,123}

\begin{tikzpicture}[
  font=\small,
  >={Stealth[length=2.6mm]},
  box/.style={draw, rounded corners=3pt, align=center, inner sep=6pt,
              minimum height=9mm},
  optbox/.style={box, draw=optpurple, fill=optpurple!8, thick},
  evalbox/.style={box, draw=evalteal, fill=evalteal!8, thick},
  fadebox/.style={box, draw=optpurple!40, fill=optpurple!3, text=black!50},
  lab/.style={font=\footnotesize, text=black!65, align=center},
  note/.style={font=\footnotesize, text=black!55, align=center},
  evalcell/.style={draw=black!55, fill=black!3, minimum width=0.5cm,
                   minimum height=0.42cm, inner sep=0pt, outer sep=0pt,
                   font=\scriptsize},
  flow/.style={->, thick, rounded corners=7pt},
  optflow/.style={flow, optpurple},
  evalflow/.style={flow, evalteal},
  fadeflow/.style={flow, optpurple!45},
  zoomline/.style={black!40, dashed, thin},
]

\node[optbox, minimum width=2.4cm] (o1) at (1.7,6.9) {optimizer 1};

\foreach \r/\y in {1/7.25, 2/6.55}{
  \node[note, anchor=east] at (4.25,\y) {\textsf{run} \r};
  \foreach \i/\c in {0/1, 1/2, 2/3, 3/4, 4/5, 5/$\cdots$, 6/$B$}
    \node[evalcell] (r\r c\i) at (4.55+0.5*\i,\y) {\c};
}
\node[evalcell, draw=optpurple, very thick, fill=optpurple!30]
      (zcell) at (5.55,6.55) {3};

\draw[decorate, decoration={brace, amplitude=4pt}] (3.2,6.32) -- (3.2,7.48);
\draw[decorate, decoration={brace, amplitude=4pt}] (4.3,7.55) -- (7.8,7.55)
      node[note, midway, above=5pt] (blab) {\textsf{budget}: $B$ \textsf{evaluations}};

\node[text=black!50, font=\Large] at (1.7,6.22) {$\vdots$};
\node[fadebox, minimum width=2.4cm] (oM) at (1.7,5.35) {optimizer $M$};

\node[optbox, minimum width=2.5cm, minimum height=1.25cm] (opt) at (1.2,0.4)
      {\textbf{optimizer}\\[-1pt] {\footnotesize black-box search}};

\node[optbox] (genome) at (3.,2.3) {genome};
\node[evalbox] (decode) at (5.85,2.3) {decode};
\node[evalbox] (inst) at (9.0,2.3) {instance};

\node[ellipse, draw=evalteal, very thick, fill=evalteal!7,
      minimum width=2.8cm, minimum height=2.0cm] (solver) at (10.75,0.4)
      {\textbf{Solver}};

\node[optbox] (F) at (6.75,-1.5) {apply $F$};
\node[optbox]  (loss) at (4.35,-1.5) {loss};

\draw[optflow] (opt.north) |- (genome.west)
      node[lab, pos=0.3, left=1pt] {\textsf{ask}};
\draw[optflow] (genome.east) -- (decode.west);
\draw[evalflow] (decode.east) -- (inst.west);
\draw[evalflow] ([yshift=4pt]inst.east) -| ([xshift=4.5pt]solver.north)
      node[lab, pos=0.65, right=2pt] {\textsf{solve} $\times~k$};
\draw[evalflow] (inst.east) -| (solver.north);
\draw[evalflow] ([yshift=-4pt]inst.east) -| ([xshift=-4.5pt]solver.north);
\draw[evalflow] ([xshift=-4pt]solver.south) |- ([yshift=4.5pt]F.east)
      node[lab, pos=0.72, above=2pt] {\textsf{runtimes} $\times~k$};
\draw[evalflow] (solver.south) |- (F.east);
\draw[evalflow] ([xshift=4pt]solver.south) |- ([yshift=-4.5pt]F.east);
\draw[optflow] (F.west) -- (loss.east);

\node[draw=black!45, dashed, rounded corners=6pt, inner sep=15pt,
      fit=(opt)(genome)(decode)(inst)(solver)(F)(loss)] (eval) {};

\node[text=black!50, font=\large] at (-0.15,-2.95) {$\cdots$};
\draw[fadeflow] (0.3,-2.95) -| (opt.south)
      node[lab, pos=0.45, below=3pt, text=black!45] {from evaluation 2};
\node[fadebox, minimum width=2.3cm] (nexteval) at (7.7,-3.15) {evaluation 4};
\draw[optflow] (loss.south) |- (nexteval.west)
      node[lab, pos=0.62, above=1pt] {\textsf{tell}};
\draw[fadeflow] (nexteval.east) -- ++(1.0,0);
\node[text=black!50, font=\large] at ($(nexteval.east)+(1.45,0)$) {$\cdots$};

\draw[zoomline] (zcell.south west) -- (eval.north west);
\draw[zoomline] (zcell.south east) -- (eval.north east);
\node[lab] at (5.9,3.7) {one evaluation \;=\; one ask/tell round};

\node[draw=black!45, dashed, rounded corners=6pt, inner sep=9pt,
      fit=(o1)(oM)(r1c6)(r2c0)(zcell)(blab)] (cfg) {};
\node[anchor=south west, font=\footnotesize\bfseries, text=black!60]
      at ($(cfg.north west)+(0,0.07)$)
      {Config:\; task,\; encoding,\; $F$};

\end{tikzpicture}}
  \caption{The evaluation pipeline. Bottom: one evaluation -- the
  optimizer asks for a genome, the encoding decodes it into an
  instance, the solver solves it $k$ times, the aggregated time is
  transformed into a loss and told back to the optimizer. Top: the
  screening stage (Stage~1) runs this loop for every optimizer with budget
  $B$ and several independent runs per configuration (task, encoding,
  loss transform~$F$).}
  \label{fig:pipeline}
\end{figure*}

\Cref{fig:pipeline} summarizes the evaluation loop and how it sits
inside the screening tournament.

\paragraph{Stage 0: calibrate the size.}
Sample random instances across a range of sizes and pick the size at
which one \emph{evaluation} of a random instance
(\cref{sec:setup}) takes roughly \calibLo--\calibHi{}\,ms. This window is a
compromise: cheap enough that an optimizer can afford tens of thousands
of evaluations, and expensive enough that timing measurements are not
dominated by noise. All hardness gains reported later are relative to
this deliberately easy starting point.

\paragraph{Stage 1: screen.}
The screen we recommend is deliberately small: a fixed roster
of \numRoster{} optimizers, each run on every encoding with a budget
of \screenBudget{} evaluations and \screenRuns{} independent runs,
under the plain loss $-t$. The roster\footnote{By Nevergrad name.
Wizards: \textsf{NGOptRW}, \textsf{NgIohMLn}, \textsf{NgIohRS}, \textsf{Shiwa}; CMA-ES family:
\textsf{MultiScaleCMA}, \textsf{MultiCMA}, \textsf{TripleCMA}, \textsf{PolyCMA}, \textsf{BFGSCMA}, \textsf{DiagonalCMA},
\textsf{MetaModelDiagonalCMA}, \textsf{ChainCMAwithMetaRecentering}; differential
evolution: \textsf{TwoPointsDE}, \textsf{BPRotationInvariantDE}; discrete:
\textsf{DiscreteBSOOnePlusOne}, \textsf{AnisotropicAdaptiveDiscreteOnePlusOne},
\textsf{RecombiningDiscreteLenglerOnePlusOne}; particle swarm: \textsf{PSO},
\textsf{RealSpacePSO}, \textsf{SQORealSpacePSO}.}, the union of the top-three
finishers across all our full screens (\appref{app:screening}), spans
the CMA-ES family, discrete $(1{+}1)$-type methods, differential
evolution, particle swarm, and the wizard portfolios, so the
practitioner never has to choose among Nevergrad's several hundred
optimizers. Optimizers are ranked by the re-measured hardness of the
best instance each finds; the outcome of the screening
is not a single winner but a ranking of optimizer \emph{families}, and
the best two or three families proceed to Stage~2.

The roster is distilled from our one-time full screen, a
meta-experiment that shaped the workflow itself. We report it here
rather than among the per-problem results of \cref{sec:results}.
We ran up to \numOptimizers{} Nevergrad
optimizers for every encoding and each of three loss transforms of the
measured time $t$: $-t$, $-\log t$, and $-t^2$. A transform reshapes
what the optimizer sees: $-\log t$ equalizes relative
gains, $-t^2$ sharpens the top.
(\numParallelOptimizers{} supported the parallel ask/tell
configuration, the rest ran serially, and a few optimizer--encoding
combinations failed.) Across all seven encoding
screens (two SAT, one TSP, three HAM, one NPFS), only \numRoster{}
distinct optimizers ever placed in the top
three of an encoding (\appref{app:screening}), and the winners
depended on the encoding: CMA-ES variants led on the
continuous duration and clause encodings, discrete $(1{+}1)$-type
methods placed high on the binary adjacency encoding, and wizard
portfolios won the TSP screen and two of the three HAM screens. The
winning optimizer, and sometimes its family, varied with the loss
transform (\appref{tab:screening}), but no transform dominated across
encodings, so the default uses the simplest, $-t$, alone. This full screen consumed roughly half of our total
compute (\cref{sec:discussion}); repeating it is warranted only when
the fixed roster fails on a new problem or encoding class.

\paragraph{Stage 2: escalate.}
Rerun the top optimizers with a budget of \escalBudget{} evaluations
and \escalRuns{} runs; the champions reported in \cref{sec:results}
come from these full-budget runs.

\paragraph{Measurement discipline.}
Every solve, during search and final measurement alike, runs under a
per-solve time cap stated in \cref{sec:setup}; an evaluation budget
therefore also bounds total compute, by budget $\times$ cap.
Two further rules proved essential. First, search and measurement are separated:
searches run massively in parallel on a shared, heterogeneous cluster
whose timings are unreliable, so every final number in the result
tables comes from a subsequent clean re-measurement (``rebench'') of the
candidate instances on a single reference machine (\cref{sec:setup});
the screening \iffullversion appendix\else table in the full version\fi{}
keeps the original screening campaign's rebench values, identified as such. Second, each evaluation aggregates repeated
measurements and, where applicable, several solver seeds. As a
default we recommend the median of at least three repeated
evaluations, aggregated across solver seeds by a second median. The
domain-specific deviations, stated in \cref{sec:setup}, trade
measurement noise against evaluation throughput and affect only
the search. For
randomized solvers, the seeds used during
breeding are kept separate from held-out seeds used only in the final
evaluation (\cref{sec:results:tsp}); optimizing against a fixed seed
set invites overfitting to those seeds, and the held-out numbers show
this effect is real.

\subsection{Case-study instantiations}
\label{sec:problems}

The four case studies emphasize different axes of the workflow: SAT
the portfolio objective, HAM the comparison of encodings, NPFS solver
specificity. The solvers are, in each domain, standard exact solvers.

\paragraph{SAT.}
Instances are 3-CNF formulas with 200 variables and 1000 clauses.
The size comes from the Stage-0 calibration, not from any prior on
hard clause--variable ratios. The mprefix encoding below lets the
search tune the ratio itself. The
basic encoding (\emph{kcnf}) is an array of $1000 \times 3$ integers,
each an index of one of the $400$ literals: the value ranges
over $[-200,199]$, negative values $-v$ decode to the positive literal
$x_v$ and non-negative values $v$ to the negated literal $\lnot
x_{v+1}$. For example, $-5$ in the second slot of clause $c$ makes $x_5$ the
second literal of $c$.
A second encoding (\emph{mprefix}) adds one extra search variable,
the \emph{prefix length} $m$: only the first $m$ clauses are active,
so the optimizer can also tune the clause--variable ratio. The solver is a
portfolio of \textsf{MiniSat}~2.2~\cite{EenS03}, \textsf{CaDiCaL} preprocessed by
\textsf{BreakID}~\cite{BiereFFH24,DevriendtBBD16}, and
\textsf{CryptoMiniSat}~\cite{SoosGM20}, run through PySAT~\cite{IgnatievMM18};
the objective is the minimum over the three. The members cover
complementary techniques (clause learning, symmetry breaking, XOR
reasoning), so an instance that merely trips one technique
while remaining easy for another does not count as hard.
Kissat~\cite{BiereFFH20}
and \textsf{march}~\cite{HeuleKWB12} are measured for context but are not part of
the objective: they serve as out-of-objective transfer checks, and the
four-solver campaign of \cref{sec:results:sat} later brings both into
the objective.

\paragraph{TSP.}
Instances are symmetric integer distance matrices on $N=20$ cities,
encoded as $\binom{20}{2}=190$ reals in $[1,U]$, $U \in
\{10^3, 10^4\}$, rounded to the nearest integer before solving; they
need not be Euclidean or metric. The solver is
\textsf{Concorde}~\cite{ApplegateBCC06}, the standard exact TSP solver.

\paragraph{Hamiltonian cycle (HAM).}
Instances are graphs on 20 vertices, in three encodings:
the 0/1 adjacency vector; per-edge scores with an edge count
$k$ (the $k$ best-scored vertex pairs become edges); and per-edge
scores with a threshold (pairs scoring below it become edges). HAM is
the one case study where we compare encodings of the same object, and
the comparison shows the encoding choice indeed matters: different optimizer families lead the
three screens (\appref{app:screening}), and the best champion comes from
the score-and-threshold encoding while the adjacency encoding's best
is weaker (\cref{tab:ham}).

Regardless of the encoding, each HAM instance is then encoded as a TSP
instance with 0/1 weights: edges get weight 0, non-edges weight 1,
so the graph is Hamiltonian exactly if the optimal tour has length 0.
The result is again solved with \textsf{Concorde}.
(Note that \textsf{Concorde} proves the exact optimum, the minimum number of absent
edges any tour must use, which subsumes the Hamiltonicity decision;
measured hardness refers to this full optimality proof.)

\paragraph{NPFS.}
We study the general, or non-permutation, flow-shop
problem~\cite{RossitTF18}, $F_m\|C_{\max}$ in the standard
three-field notation: $n$ jobs each pass through machines
$1,\dots,m$ in that fixed order, with a given duration on each; a
machine processes one job at a time, without preemption, and may
sequence the jobs in its own order; the objective is the makespan,
the completion time of the last job. It is a broader problem
than permutation flow shop ($F_m\mid\mathrm{prmu}\mid C_{\max}$),
which requires one common job order on
every machine. Instances have $6\times4$, $8\times4$, or $10\times5$
jobs${}\times{}$machines and durations in $[1,100]$, encoded as the flattened
duration matrix. We optimize against two different solvers separately:
OR-Tools \textsf{CP-SAT}~\cite{PerronDG23} and IBM
\textsf{CP~Optimizer}~\cite{LaborieRSV18} (the leading freely available and
commercial constraint-programming engines, respectively), both
single-threaded and deterministic per seed, minimizing makespan.

The baseline families for each problem are introduced with the
corresponding results in \cref{sec:results}.

\subsection{Experimental setup}
\label{sec:setup}

Searches ran on a shared university cluster (heterogeneous CPUs
spanning 2014--2021 microarchitectures, preemptible, 12-hour walltime
cap; details in \appin{app:hardware}); its timings were used only to
steer the optimizers, never for reporting. All reported times
come from one reference machine, an AMD Ryzen~5~8600G (6 cores, 12
threads, 32\,GB RAM) running Linux, capped at \timecap{}\,s in all
tables (\FSSPlargeCap{}\,s for the $10\times5$ flow-shop
experiment). Each solve ran single-threaded, pinned to one logical CPU, and
solves within a measurement campaign ran sequentially; independent
campaigns ran concurrently on distinct physical cores, and frequency
boost remained enabled. Champions, random controls, and external
baselines were measured under these same conditions;
repetition counts and seed sets are stated per experiment.

All measurements are CPU time of the solver, not wall clock, which
further insulates them from process contention effects. Solvers that run in
process (PySAT, \textsf{Concorde} via its Python bindings, \textsf{CP-SAT}) are timed
with \texttt{time.process\_time()}; \textsf{CP~Optimizer} runs as a separate
engine process and is timed as the delta in cumulative child-process
CPU time (\texttt{getrusage}). Versions: Python~3.12 with PySAT~1.9.dev2
(\textsf{MiniSat}~2.2, \textsf{CaDiCaL}~1.9.5), \textsf{CryptoMiniSat}~5.14.4,
\textsf{BreakID}~3.1, \textsf{kissat}~4.0.4, \textsf{march\_cu},
\textsf{Concorde}~03.12.19 (via pyconcorde~0.1.0),
OR-Tools~\ortoolsVersion, and \textsf{CP~Optimizer}~22.1.2. The main campaigns use master seed
\masterSeed{} for instance generation and optimizer seeding;
separately sourced auxiliary runs retain the seeds recorded in their
archived configurations. Solver seed sets are stated per experiment.

The per-evaluation aggregation depended on the solver and was chosen empirically, trading measurement noise against evaluation throughput. SAT used the
median of three repeated portfolio evaluations during the search and
the median of five sequential evaluations per solver in the final
tables. For TSP and HAM, each
candidate was solved three times at each of \textsf{Concorde} seeds 1--7; we
took a geometric mean within a seed and then the median across seeds.
The broad external scans (TSPLIB, Hard-TSPLIB, a locally generated
clustered Euclidean set, FHCP) used one solve per seed instead of three; specific instances which we refer to further (gr229, and the hardest solved
Hard-TSPLIB instance) use the full three-repetition protocol.
``Typical random'' figures are the median over 1{,}000 fresh
uniform draws from the same encoding, one evaluation per draw.
The four flow-shop arms were bred with per-arm seed rules
rather than one uniform rule ($6\times4$ \textsf{CP~Optimizer}: median over
solver seeds 1--3; $6\times4$ \textsf{CP-SAT}: seed 0 three times; $8\times4$:
one fixed seed; $10\times5$: minimum over two seeds). The
inconsistency is confined to the search, where aggregation only
steers the optimizer; every number we report comes from the uniform
final protocol below. Final tables use seeds
1--7 for remeasurement and 8--30 to test transfer.
One TSP result is from a separate $U=10^4$ experiment: 50,000 evaluations,
two runs, and the minimum over seven seeds; its final held-out median
uses seeds 8--30. The reported $6\times4$ \textsf{CP-SAT} flow-shop champion
comes from Stage~1 (\screenBudget{} evaluations, \screenRuns{} runs);
there was no Stage~2 \textsf{CP-SAT} arm. The $10\times5$ flow-shop experiment
used \FSSPlargeBudget{} evaluations and \FSSPlargeRuns{} runs. The other
full-budget runs use \escalBudget{} evaluations and \escalRuns{} runs.

The full
experiment configurations, the champion instances, and the scripts
that regenerate the final result tables are part of the artifact, archived
at \url{https://doi.org/10.5281/zenodo.23220527} and developed at
\url{https://github.com/martinkoutecky/hard-instances-on-demand-artifact}.

\section{Quantitative results}
\label{sec:results}

For each problem we compare the evolved champions with budget-matched
\textsf{RandomSearch}, the calibration baseline, and external benchmarks. Random
controls are the hardest instances found, not typical draws; each
received at least one guided run's evaluation budget (more where noted).
Budgets are matched in \emph{evaluations}, the standard budget
unit in black-box optimization. Since evaluations slow as instances
harden, this gives the guided runs more total compute than their
controls. Several controls therefore received larger pools, up to
six times the guided evaluations, flagged where they occur.
\textsf{RandomSearch} samples uniformly in the chosen
encoding, which need not induce a uniform distribution over decoded
instances. All reported times come from clean
remeasurement on the reference machine. \Cref{tab:summary} gives the primary
comparison for each case study.

\begin{table*}[t]
\scriptsize
\caption{Primary results in seconds on the reference machine. A
slash separates train from held-out; single entries have no held-out value.
The TSP control is the conservative best-held-out \textsf{RandomSearch} result.}
\smallskip
\label{tab:summary}
\centering
\begin{tabular*}{\textwidth}{@{\extracolsep{\fill}}p{0.18\textwidth}p{0.16\textwidth}p{0.18\textwidth}p{0.38\textwidth}@{}}
\toprule
Case & \shortstack[l]{Champion\\train/held-out} & \shortstack[l]{Random control\\train/held-out} & Context: external benchmarks and properties \\
\midrule
SAT portfolio & \SATevolvedTime & \SATrandomControl & phase-transition random 0.94; van der Waerden $>60$; SAT Comp: only expert-designed families harder \\
TSP, \textsf{Shiwa} $U=10^3$ & \TSPevolvedTimeB{} / \TSPevolvedHeldoutB & \TSPrandomControlTrain{} / \TSPrandomControlHeldout & gr229 \TSPlibHardest{} / \TSPlibHardestHeldout; Hard-TSPLIB $>60$ \\
HAM & \HAMevolvedTime{} / \HAMevolvedHeldout & \HAMrandomControlTrain{} / \HAMrandomControlHeldout & density-matched random $G(20,0.22)$: \HAMnullTrain{} / \HAMnullHeldout; FHCP is stronger at larger sizes \\
NPFS, \textsf{CP-Opt} $10\times5$ & \FSSPlargeChamp{} / \FSSPlargeChampHeld & \FSSPlargeRandom & proven optimal on all 30 seeds; \textsf{CP-SAT} finds no proof within \FSSPlargeCap{}\,s; hardest official VRF $10\times5$ under NPFS: \FSSPvrfHardest\,s; hardest official Taillard $20\times5$: \FSSPtaillardHardest\,s (seeds 1--7) \\
NPFS, \textsf{CP-Opt} $8\times4$ & \FSSPbigChamp{} / \FSSPbigChampHeld & \FSSPrandomControl{} / \FSSPrandomControlHeld & bred on \textsf{CP-Opt}, exceeds \textsf{CP-SAT}'s cap; Taillard starts at $20\times5$, Demirkol at $20\times15$ \\
\bottomrule
\end{tabular*}
\end{table*}

\subsection{SAT}
\label{sec:results:sat}

\paragraph{Guided search versus random control.}
The plain-encoding champion, found by \SATbestOpt{}, takes
\SATevolvedTime{}\,s for the three-solver portfolio (\cref{tab:sat}).
This is \SATvsRandomSearch$\times$ the \SATrandomControl{}\,s maximum found
by budget-matched \textsf{RandomSearch}. The prefix-encoding champion reaches
\SATevolvedTimeKnob{}\,s.
For context, a \emph{typical} formula drawn from the same encoding solves in
about \SATtypical{}\,s (median of 1{,}000 fresh draws;
\cref{sec:setup}); the champion sits roughly
\SATvsTypical$\times$ above it.

\paragraph{Clean remeasurement and transfer.}
The table reports, for each solver, the median of five sequential
evaluations. The SAT solvers do not use a randomized train/held-out split, so
there is no held-out column. Out-of-objective transfer checks come
from \textsf{kissat} and \textsf{march}: \textsf{march} solves the two champions in 0.63\,s and 0.42\,s,
respectively, showing that hardness for the three-solver objective does not
automatically extend to a fourth solver.
Hardness follows the objective, however. A separate escalation
at the same scale (\escalBudget{} evaluations, \escalRuns{} runs) bred
instances against a four-solver, paradigm-diverse portfolio: \textsf{kissat}
(CDCL), \textsf{BreakID}-preprocessed \textsf{CaDiCaL} (symmetry), \textsf{CryptoMiniSat} (XOR
reasoning), and the look-ahead solver \textsf{march}. Its champions are hard
for every paradigm at once: the best takes \SATdpllChamp{}\,s under
the minimum over the four solvers (\textsf{kissat} \SATdpllKissat{}\,s,
\textsf{BreakID}+\textsf{CaDiCaL} \SATdpllCadical{}\,s, \textsf{CryptoMiniSat} \SATdpllCMS{}\,s,
\textsf{march} \SATdpllMarch{}\,s; median of five per solver as in
\cref{tab:sat}), over \SATdpllVsTypical$\times$ a typical draw under
the same objective, and all four bred champions exceed
\SATdpllSlowest{}\,s. The absolute times are smaller than the
three-solver champion's because the objective is harsher: an
instance is credited only its \emph{fastest} paradigm. But unlike
the three-solver champions, these are hard for \textsf{march} too.

\paragraph{External context.}
The champion is \SATvsRandomPT$\times$ slower than the measured random
formula at the satisfiability phase transition. Under the portfolio minimum,
it also exceeds the pigeonhole and Tseitin instances in \cref{tab:sat}:
PHP$^{12}_{11}$ asserts that 12 pigeons fit injectively into
11 holes and requires exponential resolution
proofs~\cite{Haken85}; the Tseitin instance encodes an
unsatisfiable parity constraint over an expander graph. Each such
crafted family is easy for at least one portfolio member. The champion
does not reach the van der Waerden formula, which exceeds the cap for
every measured solver. That formula is unsatisfiable since the
van der Waerden number $W(2;3,13)$ is 160: no two-coloring of
$\{1,\dots,160\}$ avoids both length-3 monochromatic progressions in
one color and length-13 in the other. Such
formulas~\cite{KourilP08} encode substantial problem-specific
knowledge into their hardness.

We also measured the champions against the SAT Competition
archive: the 498 competition-track formulas with at most 200 variables
and 1{,}000 clauses (2002--2025) in the Global Benchmark
Database,\footnote{\url{https://benchmark-database.de/}} under
the same portfolio objective. Of these, \SATcompEasier{}
(\SATcompEasierPct\%) measured easier
than even the weaker (\SATevolvedTimeKnob{}\,s) of our two
champions.
The corpus was screened once per formula at a
12\,s cap, with representatives of the surviving families remeasured at
longer caps.
Every formula in the measured corpus that matches or exceeds
the weaker champion belongs to one of seven crafted or generator-built
families (SGen and its
balanced variant~\cite{Spence10}, hidden-weighted-bit equivalence miters,
subset-cardinality formulas, PMG, HGen, and perturbed pigeonhole
variants), which hide a global counting, equivalence, or
parity contradiction from clause-level reasoning. No application,
planning, coloring, or random-track formula of this size comes close.
In contrast, our search was given none of these semantic building
blocks.

\begin{table}[t]
\footnotesize
\caption{SAT: per-solver times in seconds on the reference
machine, cap \timecap{}\,s; each entry is the median of five sequential
evaluations. The portfolio objective is the minimum over
the first three solvers (bold); \textsf{kissat} and \textsf{march} are context only. All listed
instances are unsatisfiable. The first row (phase-transition) is one measured
random formula; the RS rows are the hardest instances found by
budget-matched \textsf{RandomSearch}. The Tseitin row is a fresh
formula of the same size class from the Global Benchmark Database.}
\label{tab:sat}
\centering
\scriptsize\setlength{\tabcolsep}{1.6pt}%
\begin{tabular}{lrrrrrrr}
\toprule
family & $n$ & $m$ & \multicolumn{1}{c}{minisat} & \multicolumn{1}{c}{BID+cdc} & \multicolumn{1}{c}{CMS} & \multicolumn{1}{c}{kissat} & \multicolumn{1}{c}{march} \\
\midrule
random, ratio 4.27      & 200 & 854  & \textbf{0.94} & 1.06 & 6.31 & 1.18 & 0.12 \\
PHP$^{12}_{11}$         & 132 & 738  & $>$60 & \textbf{0.009} & $>$60 & 36.2 & 59.6 \\
Tseitin expander        & 150 & 400  & $>$60 & \textbf{0.008} & 0.80 & 10.4 & $>$60 \\
$W(2;3,13)$             & 160 & 7308 & $\boldsymbol{>}$\textbf{60} & $\boldsymbol{>}$\textbf{60} & $\boldsymbol{>}$\textbf{60} & $>$60 & $>$60 \\
RS (kcnf)               & 200 & 1000 & \textbf{0.07} & 0.10 & 0.16 & 0.19 & 0.02 \\
RS (mprefix)            & 200 & 852  & \textbf{0.47} & 0.62 & 2.45 & 0.59 & 0.08 \\
champion (kcnf)         & 200 & 1000 & \textbf{12.0} & 17.8 & $>$60 & 9.2 & 0.63 \\
champion (mprefix)      & 200 & 889  & \textbf{11.1} & 13.6 & 53.4 & 10.2 & 0.42 \\
\bottomrule
\end{tabular}
\end{table}

\subsection{TSP}
\label{sec:results:tsp}

\paragraph{Guided search versus random control.}
The headline TSP result is the \textsf{Shiwa} champion at $U=10^3$: it takes
\TSPevolvedTimeB{}\,s on the breeding seeds and
\TSPevolvedHeldoutB{}\,s held-out (\cref{tab:tsp}), compared with
\TSPrandomControlTrain{}\,s and \TSPrandomControlHeldout{}\,s for the
conservative best-held-out \textsf{RandomSearch} control. In an auxiliary $U=10^4$
experiment (protocol in \cref{sec:setup}), \textsf{DiagonalCMA} reached
\TSPevolvedTime{}\,s train and \TSPevolvedHeldout{}\,s held-out.

\paragraph{Clean remeasurement and held-out transfer.}
\textsf{Concorde} is randomized. The main experiment reports the median over breeding
seeds 1--7 and, separately, over 23 unseen seeds 8--30. This separation
exposes seed overfitting: the plain \textsf{CMA} champion falls from 9.97\,s on
its breeding seeds to 1.31\,s held-out, a $7.6\times$ drop. The
\textsf{RandomSearch} control row instead uses the candidate with the best
held-out value (\TSPrandomControlTrain{}/\TSPrandomControlHeldout{}\,s),
a deliberately conservative choice.

\paragraph{External context.}
On the held-out seeds, gr229, the hardest of the 58 classical
TSPLIB instances with at most 300 cities that we measured, takes
\TSPlibHardestHeldout{}\,s. \textsf{Shiwa} is therefore
\TSPShiwaVsTSPLibHeldout$\times$ harder despite having 20 cities to
gr229's 229; the auxiliary \textsf{DiagonalCMA} result is
\TSPDiagVsTSPLibHeldout$\times$ harder. Neither champion reaches the
hand-crafted Hard-TSPLIB instances~\cite{VercesiEtAl23}, which target
\textsf{Concorde}'s branch-and-cut following Hougardy and Zhong~\cite{HougardyZ21};
most measured instances in that set exceed the cap.

\begin{table}[t]
\footnotesize
\caption{TSP: median \textsf{Concorde} process time in seconds on the reference
machine, cap \timecap{}\,s. Train = breeding seeds 1--7; held-out =
unseen seeds 8--30. Generated instances and the two fixed comparators
(gr229, hardest solved Hard-TSPLIB) use three repetitions per seed; the
broad TSPLIB, clustered, and Hard-TSPLIB scans use one solve per seed.
Random draws use budget $100{,}000$, matching the $U=1000$ setup and
twice the \textsf{DiagonalCMA} budget.}
\label{tab:tsp}
\centering
\scriptsize\setlength{\tabcolsep}{2pt}%
\begin{tabular}{lrrr}
\toprule
family & $N$ & train & held-out \\
\midrule
random Euclidean, hardest of draw & 20   & 0.08 & 0.07 \\
TSPLIB, median                & $\le$300 & 0.097 & -- \\
TSPLIB, hardest (gr229)       & 229     & 2.34  & 2.24 \\
clustered Euclidean (our generator), median   & 50--200 & 0.17 & -- \\
Hard-TSPLIB, hardest solved   & 20      & 33.8  & 28.7 \\
Hard-TSPLIB (23 of 40)        & 10--76  & $>$60 & -- \\
\textsf{RandomSearch}, best find       & 20      & 0.85  & 0.75 \\
\textsf{RandomSearch}, best held-out   & 20      & \TSPrandomControlTrain & \TSPrandomControlHeldout \\
champion \textsf{CMA}                  & 20      & 9.97  & 1.31 \\
champion \textsf{Shiwa}                & 20      & \TSPevolvedTimeB  & \TSPevolvedHeldoutB \\
champion \TSPbestOpt          & 20      & \textbf{\TSPevolvedTime} & \textbf{\TSPevolvedHeldout} \\
\bottomrule
\end{tabular}
\end{table}

\subsection{Hamiltonian cycle}
\label{sec:results:ham}

\paragraph{Guided search versus random control.}
The \HAMbestOpt{} champion takes \HAMevolvedTime{}\,s on the breeding
seeds and \HAMevolvedHeldout{}\,s held-out (\cref{tab:ham}). The
strongest instance from three fresh budget-matched \textsf{RandomSearch}
campaigns, one per encoding, takes \HAMrandomControlTrain{}\,s
and \HAMrandomControlHeldout{}\,s, so the guided gain is
\HAMvsRandomTrain$\times$ on
training seeds and \HAMvsRandomHeldout$\times$ held-out.

\paragraph{Clean remeasurement and held-out transfer.}
The table separates breeding seeds 1--7 from unseen seeds 8--30.
A random instance near the density of our champion (drawn from $G(20,0.22)$) reaches \HAMnullTrain{}\,s train
and \HAMnullHeldout{}\,s held-out.
This baseline is deliberately strong: its edge probability
sits near both the champions' density and the $n=20$ Hamiltonicity
threshold, insights that would be unknown for a new problem.

\paragraph{External context.}
There is no curated hard family at $N=20$. The FHCP Challenge
Set~\cite{Haythorpe18} starts at $N=66$; its hardest member that \textsf{Concorde}
solves on all seven measured seeds takes \HAMfhcpSolved{}\,s at $N=108$,
and six of the eleven scanned members exceed the cap. FHCP is therefore stronger, but only at larger sizes. The closest size-matched
precedent is not a curated library but evolved instances: Sleegers and
Van den Berg's hardest 14-vertex graphs~\cite{SleegersB22}, bred against
a backtracking solver rather than \textsf{Concorde}.

\begin{table}[t]
\footnotesize
\caption{HAM: median \textsf{Concorde} process time in seconds, cap
\timecap{}\,s. Train = breeding seeds 1--7, held-out = seeds
8--30. Generated instances use three repetitions per seed;
the FHCP scan uses one solve per seed.}
\label{tab:ham}
\centering
\scriptsize\setlength{\tabcolsep}{2pt}%
\begin{tabular}{lrrr}
\toprule
family & $N$ & train & held-out \\
\midrule
random $G(20,0.22)$ & 20 & \HAMnullTrain & \HAMnullHeldout \\
random cubic & 20 & 0.04 & 0.01 \\
FHCP, hardest solved on all seeds & 108 & \HAMfhcpSolved & -- \\
FHCP, 6 of 11 scanned members & 90--150 & $>$60 & -- \\
\textsf{RandomSearch}, best of three encodings & 20 & \HAMrandomControlTrain & \HAMrandomControlHeldout \\
champion \textsf{TwoPointsDE} (scores + threshold) & 20 & 0.61 & 0.27 \\
champion \textsf{NGOptRW} (adjacency) & 20 & 0.86 & 0.32 \\
champion \HAMbestOpt{} (scores + threshold) & 20 & \textbf{\HAMevolvedTime} & \textbf{\HAMevolvedHeldout} \\
\bottomrule
\end{tabular}
\end{table}

\subsection{Non-permutation flow-shop scheduling}
\label{sec:results:jssp}

\paragraph{Guided search versus random control.}
The strongest non-permutation flow-shop result is the $10\times5$
\textsf{CP~Optimizer} champion, bred by differential evolution (\textsf{DE}): it takes
\FSSPlargeChamp{}\,s on
seeds 1--7 and \FSSPlargeChampHeld{}\,s held-out (\cref{tab:fssp}),
about a factor of \FSSPlargeVsRandom{} above the \FSSPlargeRandom{}\,s maximum
among 100,000 uniform random draws. The comparison is conservative, since
the random draw received twice the evaluations of each guided run. At
$8\times4$, the \FSSPbestOpt{} champion takes \FSSPbigChamp{}\,s on
evaluation seeds 1--7 and \FSSPbigChampHeld{}\,s held-out, a factor of
\FSSPbigVsHardestRandom{} above the hardest of its own 100,000
budget-matched draws (\FSSPrandomControl{}\,s).
The median of
1{,}000 fresh uniform draws is \FSSPbigTypical{}\,s at $8\times4$ and
\FSSPlargeTypical{}\,s at $10\times5$. The champions sit factors of
\FSSPbigVsTypical{} and \FSSPlargeVsTypical{} above their
typical draw.

\paragraph{Clean remeasurement and held-out transfer.}
The $10\times5$ champion was proved
optimal on all 30 \textsf{CP~Optimizer} seeds; \textsf{CP-SAT} found a feasible schedule
but no optimality proof within the \FSSPlargeCap{}\,s cap. The
$8\times4$ champion exceeds the \timecap{}\,s cap on \textsf{CP-SAT} for every
tested seed. Both \textsf{CP-Optimizer}-bred champions thus remain hard for the
engine they were never bred against, in contrast to the weak transfer
at $6\times4$ below.

\paragraph{External context.}
Established NPFS experiments use the 40 Demirkol flow-shop
makespan instances~\cite{DemirkolMU98}, whose smallest group is
$20\times15$, and re-solve the Taillard and VRF processing-time
matrices under NPFS~\cite{BenavidesR16,BenavidesR18}. Taillard~\cite{Taillard93}
and VRF~\cite{ValladaRF15} were introduced for the permutation
variant, so we recompute their difficulty for our model rather than
quote published results. VRF contains ten exact $10\times5$ matrices,
size-matching our champion; Taillard's smallest native group is
$20\times5$. From both suites we re-solved the size-closest instances
under NPFS with \textsf{CP~Optimizer} on the reference machine; every
solve reached proven optimality on all ten solver seeds. Over the champion's evaluation seeds 1--7,
the hardest VRF matrix (\texttt{VFR10\_5\_2}) has a median of
\FSSPvrfHardest{}\,s and the hardest Taillard matrix (\texttt{ta005})
\FSSPtaillardHardest{}\,s. Our champion takes
\FSSPlargeVsVRF$\times$ longer than the former at identical size, and
\FSSPlargeVsTaillard$\times$ longer than the latter with half the
jobs.

\paragraph{The $6\times4$ solver-specificity arms.}
The \textsf{CP-SAT} champion is a Stage~1 experiment: it used
\screenBudget{} evaluations rather than the full Stage~2 budget. It takes
\FSSPsatChamp{}\,s on evaluation seeds 1--7 and
\FSSPsatChampHeld{}\,s held-out, versus 0.21/0.25\,s for a \textsf{RandomSearch}
control drawn with 100,000 evaluations. That random comparison is
conservative because it uses more than six times the per-run search
budget. The
champion transfers weakly to \textsf{CP~Optimizer} at \FSSPsatChampTransfer{}\,s.
Conversely, the
$6\times4$ \textsf{CP-Optimizer} champion reaches only
\FSSPoptChamp{}/\FSSPoptChampHeld{}\,s, compared with 0.18/0.18\,s for
its \textsf{RandomSearch} control.

\begin{table}[t]
\footnotesize
\caption{Non-permutation flow shop: median process time in seconds per engine, cap
\timecap{}\,s except \FSSPlargeCap{}\,s for $10\times5$. Engine columns use
evaluation seeds 1--7; the last
column uses the champion's own engine on seeds 8--30.
Random rows are the hardest of $100{,}000$ fresh draws
(for the \textsf{CP-SAT} arm this exceeds the search budget, a
conservative comparison). In the $>$-entries, \textsf{CP-SAT} found a feasible
schedule but no optimality proof within the cap.}
\label{tab:fssp}
\centering
\scriptsize\setlength{\tabcolsep}{3pt}%
\begin{tabular}{lrrrr}
\toprule
family & $n{\times}m$ & \textsf{CP-SAT} & \textsf{CP-Opt} & held-out \\
\midrule
champion, \textsf{CP-SAT} arm & $6\times4$ & \textbf{\FSSPsatChamp} & \FSSPsatChampTransfer & \FSSPsatChampHeld \\
\quad \textsf{RandomSearch}, same arm & $6\times4$ & 0.21 & 0.02 & 0.25 \\
champion, \textsf{CP-Opt} arm & $6\times4$ & 0.78 & \FSSPoptChamp & \FSSPoptChampHeld \\
\quad \textsf{RandomSearch}, same arm & $6\times4$ & 0.14 & 0.18 & 0.18 \\
\midrule
\textsf{RandomSearch}, uniform & $8\times4$ & -- & \FSSPrandomControl & 0.65 \\
champion (\FSSPbestOpt{}) & $8\times4$ & $>$60 & \textbf{\FSSPbigChamp} & \FSSPbigChampHeld \\
\midrule
\textsf{RandomSearch}, uniform & $10\times5$ & -- & \FSSPlargeRandom & -- \\
champion (\textsf{DE}) & $10\times5$ & $>$\FSSPlargeCap & \textbf{\FSSPlargeChamp} & \FSSPlargeChampHeld \\
\bottomrule
\end{tabular}
\end{table}

\section{Structural observations}
\label{sec:structure}

The search objective contains no explicit structural features,
so whatever structure the champions have was discovered through
running time within the chosen encodings, not prescribed by the
objective.
We address two questions: whether the hardness is an encoding
artifact, a degenerate instance the solver merely mishandles, and
whether the champions resemble known hard families for their
problem.
This analysis is an illustrative companion, not part of our proposed
default for experimental papers: what transfers across problems is the
pair of questions, while the features answering them are necessarily
domain-specific.

\paragraph{SAT.}
On degeneracy: both champions are syntactically clean. All
200 variables occur, every clause has width three, there are no
duplicate clauses or repeated literals, and tautologies account for
about 1\% of the clauses; the variable-degree spread matches a uniform
random formula of the same size, ruling out a degree-imbalance
explanation though not correlated substructure.
Nor can the hardness be explained merely by rediscovering the phase transition: the prefix
champion's clause--variable ratio of $4.4$ does sit just above the
random 3-SAT threshold near $4.27$, where random hardness is known
to peak~\cite{CheesemanKT91}. But a measured random formula at
the threshold solves \SATvsRandomPT$\times$ faster (\cref{tab:sat}):
the champions occupy the hard region, yet are far from typical
draws within it. Against the crafted families,
one difference is measurable: \textsf{BreakID}'s symmetry breaking and
\textsf{CryptoMiniSat}'s XOR reasoning each crack pigeonhole or Tseitin
formulas, but neither improves on plain \textsf{MiniSat} for the champions.
The evolved formulas do not expose the symmetry or XOR
structure those techniques exploit, which is plausibly why they
survive the portfolio minimum where certain crafted families do not.

\paragraph{TSP.}
An exact solver's work splits into finding an optimal tour and
proving that no better one exists. To see which part the champion
stresses, we ran 2-opt, the classical local search that swaps pairs of tour
edges until no exchange improves~\cite{Croes58}, from 300 seeded
restarts: 262 finish within
1\% of the exact optimum, so \emph{finding} near-optimal tours is
easy, in line with earlier observations~\cite{Zhang04}. The champion's distance matrix is compatible with that: nearly
uniform weights (coefficient of variation $0.08$) with no
triangle-inequality violations. A natural hypothesis,
motivated though not established by these observations, is that
many near-tied tours prolong \textsf{Concorde}'s optimality proof, echoing the
fine structure of near-optimal solutions observed in other search
spaces~\cite{GomesS99}. But this signature is not the only road to
hardness: the harder $U=10^4$ \textsf{DiagonalCMA} champion has the opposite
profile (coefficient of variation $0.35$, some triangle-inequality
violations, 86 of 300 2-opt restarts within 1\%), so the search found two
structurally different ways to slow the same solver.

\paragraph{Hamiltonian cycle.}
Here the natural question is where the champions sit relative
to the Hamiltonicity boundary.
The champions are sparse (edge density $0.19$, compared with
$0.49$ for an unconstrained random graph), asymmetric, and barely
non-Hamiltonian: the
optimal tours of the TSP reduction use only two or three absent edges.
Thus the measured instances sit
just beyond the Hamiltonicity boundary, and proving non-Hamiltonicity
near that boundary is one hypothesis for their difficulty. As with
SAT, the boundary explains only part of the hardness: a random graph
near the champions' density, $G(20,0.22)$, is $4.3\times$ (train)
and $2.8\times$ (held-out) easier (\cref{tab:ham}). This structure
differs from the high symmetry and few Hamiltonian cycles associated
with hard FHCP instances~\cite{Haythorpe18}; at $n=20$, our
search did not produce that pattern.

\paragraph{Non-permutation flow-shop scheduling.}
The quantitative results showed strong solver specificity at
$6\times4$ (\cref{tab:fssp}); the duration matrices make that
specificity visible. Both $6\times4$ champions concentrate $1.7\times$ the mean
job length in one job; they differ in machine balance, with the least
loaded machine carrying $0.4\times$ the mean load (\textsf{CP-Optimizer}-hard)
versus $0.8\times$ (\textsf{CP-SAT}-hard), and in duration range (2--99
versus 24--95). The larger \textsf{CP-Optimizer} champions share the
machine-imbalance signature: the
$8\times4$ champion has one job at $1.9\times$ the mean job length and
one machine loaded at a third of the mean, and the $10\times5$
champion has one job at $1.8\times$ the mean and one machine at half
the mean.

\section{Discussion}
\label{sec:discussion}

\paragraph{What the method does and does not deliver.}
The workflow establishes an empirical lower bound on how badly a named
solver can behave within a stated encoding and search budget. Comparisons
with budget-matched random search show what guided optimization added;
held-out seeds and clean remeasurement show how much of that gain survives
the search environment. The experiments do not establish worst-case
complexity, hardness for solvers outside the objective, or a causal account
of why a champion is slow. They also cannot distinguish whether instance
size, encoding, or search budget limits a modest gain, as in the $n=20$ HAM
study. Whether the approach remains affordable at the larger
instance sizes common in algorithm engineering is likewise open:
evaluation cost grows with the very solve times the search drives
up.

\paragraph{Hardness is solver-specific, by design.}
The SAT champions that stall CDCL solvers fall in under a second to a
DPLL-style solver; the non-permutation flow-shop \textsf{CP-SAT} champion
collapses from \FSSPsatChamp{}\,s
to \FSSPsatChampTransfer{}\,s on \textsf{CP~Optimizer}.
We report this specificity explicitly: the
method answers ``where does \emph{this} solver break.'' When broader
hardness is wanted, a portfolio objective
(minimum over several solvers) extends it to the portfolio's
members; our SAT experiments validate this. Separately, the
$8\times4$ and $10\times5$ non-permutation flow-shop champions, bred
against \textsf{CP~Optimizer} alone, are also unsolved by \textsf{CP-SAT} within their
caps.

\paragraph{Overfitting and measurement.}
Two failure modes surfaced repeatedly and shaped the protocol. First,
instances overfit to the random seeds used during breeding; in the
held-out columns of \cref{tab:tsp}, the apparent hardness of the plain
\textsf{CMA} champion shrinks by $7.6\times$, and in an earlier campaign a
random-search candidate lost essentially all of its apparent hardness
under held-out seeds. Reporting held-out numbers
should be standard for randomized solvers. Second, timing on shared
hardware is untrustworthy; instances that seemed hard under cluster
contention often were not. We therefore report only clean sequential
remeasurements, of random winners and champions alike.

\paragraph{Cost.}
The four case studies together consumed about $184{,}000$
CPU-hours of search ($25{,}000$ SAT, $81{,}000$ TSP, $49{,}000$ HAM,
$29{,}000$ flow shop; $86{,}000$ of it screening, $98{,}000$
escalation). Most of this is the price of \emph{validating the
method}: a \numOptimizers-optimizer screen across all encodings and
three loss transforms. The reduced protocol we recommend (the fixed
\numRoster-optimizer roster under the plain loss of \cref{sec:protocol},
then \escalRuns{} escalation runs of \escalBudget{} evaluations for the
two or three best families) costs a few hundred to a few thousand CPU-hours
per problem, depending on how expensive evaluations become as the
search progresses. These hours are also embarrassingly parallel:
distinct optimizers and their independent runs can all execute
concurrently, so on a many-core machine the wall-clock time approaches
one run's length -- days, not CPU-years.

\section{Future work}
\label{sec:conclusion}

Four directions seem most promising to us. \emph{Deployment
where benchmarks are thinnest:} the problems that need adversarial
suites most have no curated instances at all. The electric
fleet-size-and-mix problem of \cref{sec:intro} is a natural first
target. \emph{Scaling up:} the hardness found grows steeply with
instance size (flow shop: a factor of 33 from $8\times4$ to $10\times5$);
can small champions warm-start searches at sizes where direct search
is unaffordable? The lift itself is the obstacle: a barely
non-Hamiltonian graph, say, must be grown while remaining barely
non-Hamiltonian. \emph{Generators instead of instances:} evolving
the parameters of an instance \emph{distribution} until its typical
draw is hard would yield transferable, size-scalable families, as
suggested for perfect rectangle packing by Braam and Van den
Berg~\cite{BraamB22}. This would connect the workflow to automated instance
generation and algorithm
configuration~\cite{DangEtAl22,LopezIbanezEtAl16} and to instance
space analysis~\cite{SmithMilesL12,MunozSM20}, though a
parameterization whose typical draw is hard need not exist. \emph{Structural
objectives:} the same search can target structure directly, e.g.\
maximizing the integrality gap of the subtour relaxation on small TSP
instances, complementing exact computations~\cite{CookHP26}.

\bibliographystyle{siamplain}
\bibliography{bibliography}

\iffullversion
\appendix

\section{Screening results}
\label{app:screening}

\Cref{tab:screening} shows, per encoding, the three best optimizers of
the screening stage and their clean-rebench hardness under each loss
transform (best instance per cell). The union of these
top-three finishers is the fixed \numRoster-optimizer roster
recommended in \cref{sec:protocol}. The historical Stage~1
remeasurement checkpoints were not retained, so this table records the
recovered summary; it is the one table the artifact scripts do not
regenerate from scratch.

\begin{table}[tbhp]
\footnotesize
\caption{Screening (budget \screenBudget{}, \screenRuns{} runs): best
instance hardness in seconds after the screening campaign's own clean
rebench (not the final reference machine of \cref{app:hardware}), per
loss transform. Only rankings and relative magnitudes are used from
this table.
Three best optimizers per encoding; ``--'' = combination not sampled.}
\label{tab:screening}
\begin{center}
\scriptsize\setlength{\tabcolsep}{3pt}%
\begin{tabular}{lrrr}
\toprule
encoding / optimizer & $-t$ & $-\log t$ & $-t^2$ \\
\midrule
\multicolumn{4}{l}{HAM, adjacency matrix (triv)} \\
\quad \textsf{NGOptRW} & 1.14 & 0.82 & -- \\
\quad \textsf{NgIohMLn} & 0.83 & 1.05 & 0.63 \\
\quad \textsf{RecombiningDiscreteLenglerOnePlusOne} & 0.56 & 0.99 & 0.80 \\
\multicolumn{4}{l}{HAM, edge scores + edge count (pint)} \\
\quad \textsf{NgIohRS} & 0.89 & 0.87 & 0.86 \\
\quad \textsf{BFGSCMA} & 0.26 & 0.81 & 0.24 \\
\quad \textsf{RealSpacePSO} & -- & 0.76 & 0.41 \\
\multicolumn{4}{l}{HAM, edge scores + threshold (pthr)} \\
\quad \textsf{BPRotationInvariantDE} & 0.39 & 0.42 & 1.97 \\
\quad \textsf{DiscreteBSOOnePlusOne} & 0.92 & -- & 1.81 \\
\quad \textsf{TwoPointsDE} & 0.24 & 1.08 & -- \\
\multicolumn{4}{l}{TSP, distance matrix (dm)} \\
\quad \textsf{NgIohRS} & 15.88 & 12.88 & 5.06 \\
\quad \textsf{Shiwa} & 4.81 & 14.23 & 7.42 \\
\quad \textsf{PolyCMA} & 3.63 & 8.58 & 12.89 \\
\multicolumn{4}{l}{SAT, plain (kcnf)} \\
\quad \textsf{MultiScaleCMA} & 3.19 & 2.19 & 2.19 \\
\quad \textsf{MultiCMA} & 2.25 & 2.00 & -- \\
\quad \textsf{AnisotropicAdaptiveDiscreteOnePlusOne} & 0.57 & 0.55 & 0.58 \\
\multicolumn{4}{l}{SAT, clause prefix (mprefix)} \\
\quad \textsf{TripleCMA} & 4.77 & 5.01 & -- \\
\quad \textsf{SQORealSpacePSO} & 2.30 & 2.21 & 3.53 \\
\quad \textsf{PSO} & 2.54 & 3.45 & 2.72 \\
\multicolumn{4}{l}{NPFS, duration matrix (mat)} \\
\quad \textsf{ChainCMAwithMetaRecentering} & 0.40 & 0.43 & 0.50 \\
\quad \textsf{DiagonalCMA} & 0.44 & 0.50 & 0.40 \\
\quad \textsf{MetaModelDiagonalCMA} & 0.47 & 0.49 & 0.42 \\
\bottomrule
\end{tabular}
\end{center}
\end{table}

\section{Hardware}
\label{app:hardware}

The search cluster is a shared university HPC system (SLURM, a
preemptible partition with a 12-hour walltime cap). Its nodes span CPU
microarchitectures from Haswell (2014) to Zen~3 (2021) (Intel
Xeon E5~v3/v4 and Gold~6240; AMD EPYC 7302/7451/7532/7543), with jobs
placed on any node without a CPU-feature constraint. Because the
partition is shared and heterogeneous, raw search-time measurements
are contention-inflated and not comparable across nodes; the cluster
was used only to generate candidate instances, never as a timing
authority.

The reference machine for all final reported numbers is an AMD
Ryzen~5~8600G (6 cores/12 threads) with 32\,GB RAM running Linux.
Every measured solve was single-threaded and pinned to one logical
CPU; solves within a campaign ran sequentially, while independent
campaigns ran concurrently on distinct physical cores. Frequency boost
remained enabled throughout; CPU-frequency and temperature logs are
included in the artifact. All reported times are CPU time of the
solver process (\cref{sec:setup}). The screening table in
\cref{app:screening} predates this campaign and reports the original
screening campaign's clean-rebench values.

\fi

\end{document}